\documentstyle[aps,prl,epsfig,multicol]{revtex}

\begin{document}
\title{Synthesis and pinning properties of the infinite-layer superconductor Sr$%
_{0.9}$La$_{0.1}$CuO$_{2}$ }
\author{C. U. Jung,\cite{email} J. Y. Kim, Mun-Seog Kim, Min-Seok Park, Heon-Jung
Kim, Yushu Yao, S. Y. Lee, and Sung-Ik Lee}
\address{National Creative Research Initiative Center for Superconductivity\\
and Department of Physics, Pohang University of Science and\\
Technology, Pohang 790-784, Republic of Korea }
\maketitle

\begin{abstract}
We report the high-pressure synthesis of the electron-doped infinite-layer
superconductor Sr$_{0.9}$La$_{0.1}$CuO$_{2}$ and its superconducting
properties. A Rietveld analysis of X-ray powder diffraction data showed
that, within the resolution of the measurement, the sample had purely an
infinite-layer structure without any discernible impurities. The
superconducting volume fraction and the transition width were greatly
improved compared to those in previous reports. The irreversibility field
line and the intragranular critical current density were much higher than
those of La$_{1.85}$Sr$_{0.15}$CuO$_{4}$ and Nd$_{1.85}$Ce$_{0.15}$CuO$_{4}$%
. The stronger pinning behaviors are consistent with the strong interlayer
coupling due to the short distance between CuO$_{2}$ planes.
\end{abstract}

\begin{multicols}{2}
\section{Introduction}

The electron-doped infinite-layer compounds (Sr$_{1-x}^{+2}$Ln$_{x}^{+3}$)CuO%
$_{2}$ (Ln = La, Sm, Nd, Gd, {\it etc}.) consist of an infinite stacking of
CuO$_{2}$ planes and metallic (Sr) layers\cite{Siegrist88,Smith91}. The
charge reservoir block common to other cuprate superconductors does not
exist in these compounds. Since their structure is so simple, these
compounds provide a unique opportunity to explore the fundamental nature of
high-temperature superconductor.

Although electron-doped infinite-layer compounds have existed for quite a
while, not many studies of their properties have been done because of the
lack of high-quality bulk samples. In the case of films of these compounds,
the superconducting transition temperatures, $T_{c}$, are reported to be
much lower than those of bulk samples. For example, the $T_{c}$ of a (Sr$%
_{1-x}$Nd$_{x}$)CuO$_{2}$ film is reduced by about one half compared to that
of the bulk\cite{Sugii94,Jones94}. High-pressure synthesis is known to be a
unique method that stabilizes the bulk form of infinite-layer compounds with
larger superconducting volume fractions\cite
{Ikeda93,Jorgensen93,Er94,Er92,Er91,Review,Yao94,Er97}.

Nonetheless, several interesting properties have been observed for these
compounds. Since the distance between CuO$_{2}$ planes is short (3.41 \AA )
due to the absence of the charge reservoir block, the interlayer coupling,
and thus the superconductivity is expected to be quite strong. However, the $%
T_{c}$ is only about 43 K,\cite
{Ikeda93,Jorgensen93,Er94,Er92,Er91,Review,Yao94,Er97} and neither the ionic
radius, the magnetic moment, nor the concentration of Ln ions at Sr sites
affects $T_{c}$\cite{Ikeda93}. Moreover, the oxygen has been found to be
very stoichiometric; neither vacancies nor interstitial oxygens exist\cite
{Jorgensen93}.

Pinning is another measure of the strength of interlayer coupling. It has
been reported that pinning is enhanced by reducing the thickness of the
charge reservoir block\cite{Kim91}. However, the pinning properties of
infinite-layer superconductors have not been well studied. Strangely enough,
the reported irreversibility field line, $H_{irr}$, of Sr$_{0.9}$La$_{0.1}$%
CuO$_{2}$\ (La-112) is more than two times higher than that of La$_{1.85}$Sr$%
_{0.15}$CuO$_{4}$ (La-214) while the intragranular critical current density $%
J_{c}$ is smaller than that of La-214\cite{Er94,Senoussi87}.

In this research, we used high-quality samples to study the pinning
properties of the infinite-layer superconductor La-112.\ The superconducting
volume fraction and the transition width were determined from the low-field
magnetization data and were found to be improved over previously reported
values and a Rietveld analysis of the X-ray powder diffraction data
confirmed that the samples were of high quality. The pinning properties were
studied by measuring the irreversibility field line and the intragranular
critical current density. Contrary to a previous report,\cite{Er94} both
values were higher than those for La-214 and Nd$_{1.85}$Ce$_{0.15}$CuO$_{4}$
(Nd-214), supporting the stronger interlayer coupling between CuO$_{2}$
planes due to the absence of the charge reservoir block.

\section{experimentals}

A cubic multi-anvil-type press was used to synthesize La-112\cite{Ikeda93},
The precursors were prepared by using the solid-state reaction method\cite
{Ikeda93,Er91}, Starting materials of La$_{2}$O$_{3}$, SrCO$_{3}$, and CuO
were mixed to the nominal composition of Sr$_{0.9}$La$_{0.1}$CuO$_{2}$. The
mixture was then calcined at 950 $^{\circ }$C\ for 36 hours with several
intermittent grindings. The pelletized precursors sandwiched by Ti oxygen
getters were put into a Au capsule in a high-pressure cell. A {\it D}-type
thermocouple was used to monitor the temperature.

The pressure cell was compressed up to 4 GPa and then heat-treated using a
graphite-sleeve heater. The temperature of the Au capsule was calibrated to
the heating power, which allowed us to use the heating power to control the
temperature. However, much of the power from the power supply was lost to
the stray resistance ({\it R}$_{{\rm stray}}\sim 10^{-2}$ $\Omega $ )
between the power supply and the graphite heater ({\it R}$_{{\rm heater}%
}\sim 10^{-2}$ $\Omega $ ). Even though the power was supplied at a constant
rate by the power supply, the actual heating power of the sample fluctuated
because {\it R}$_{{\rm heater}}$ changed during the synthesis; $\Delta ${\it %
R}$_{{\rm heater}}/${\it R}$_{{\rm heater}}\sim 0.1$. The amount of
fluctuation was roughly proportional to {\it R}$_{{\rm stray}}$/{\it R}$_{%
{\rm heater}}$. To solve this problem, we controlled the heating power
across the sample, instead of the main power. With this method, a
temperature stability of $\pm $ 2 $^{\circ }$C was obtained for a two-hour
heating time under high-pressure conditions.

The heating power was increased linearly and then maintained constant for 2
hours. The synthesizing temperature was about $1000$ $^{\circ }$C. Then, the
sample was quenched to room temperature after an additional postannealing at 
$500\sim 600$ $^{\circ }$C for 4 hours. Two conditions were important in
obtaining higher quality samples. One was the long-term stability of the
synthesizing temperature, and the other was the uniformity of the
temperature inside the sample cell, the former being more important. The
pressure cell and the heating conditions were optimized after hundreds of
trials, and very homogeneous samples larger than $200$ mg were obtained. The
size of the as-grown polycrystalline samples was about 4.5 mm in diameter
and 2.8 mm in height.

The structure and the grain shape and size were studied by using powder
X-ray diffraction (XRD) measurements with Cu {\it K}$\alpha $ radiation, as
well as scanning electron microscopy (SEM) and optical microscopy. The
structural characterization at room temperature was carried out by using the
Rietveld refinement method to analyze the X-ray powder diffraction data. The
SEM image showed closely packed grains of uniform size with an average
radius $R\lesssim 5~\mu $m. To investigate the superconducting properties,
we used a SQUID magnetometer (MPMS{\it XL}, Quantum Design) to measure
magnetization curves $M(T)$ and $M(H)$. We report our study on the three
good samples which are denoted by sample A, sample B, and sample C
respectively. While all kinds of data was shown for sample A, only low-field
magnetization data was shown for the other two samples.

\section{data and discussion}

\subsection{Structure}

The Rietveld refinement profile with the tetragonal space group {\it P}4/mmm
of our sample is shown in Fig. \ref{XRD}. The value of the diffraction angle 
$2\theta $ was varied from 20$^{\circ }$ to 140$^{\circ }$ in steps of 0.02$%
^{\circ }$, and the integration time was 15 seconds at each point. The
Rietveld refinement program RIETAN-94 with 50 parameters was used for the
analysis\cite{Izumi}. In that analysis, the thermal factors were assumed to
be isotropic, and the coordination of each atom was fixed. We constrained
the Sr : La ratio to the nominal stoichiometry of the precursor\cite
{Jorgensen93}. The values obtained for the lattice constants, $a=b=3.950$
\AA\ and $c=3.410$ \AA ,\ agree quite well, within 0.001 \AA , with those
obtained using neutron powder diffraction\cite{Jorgensen93}. Within the
resolution of this refinement, no discernible amounts of impurities were
observed. The agreement factors, $R$, between the measured and the
calculated diffraction intensities were quite small, and the goodness of fit
was excellent ($S=4.0008$). The refined structural parameters are summarized
in Table \ref{refinement}.

A structural analysis of an infinite-layer compound can also give valuable
information about the doping concentration because the lattice constants are
sensitive to the doping concentration. The lattice constants are known to
behave in opposite ways with increased doping; the {\it a-}axis expands
while the {\it c}-axis shrinks\cite{Ikeda93}. The Rietveld refinement showed
that the doping concentration in our Sr$_{1-x}$La$_{x}$CuO$_{2}$ was
approximately $x=0.1$, which was the same as the nominal composition.

\subsection{Superconducting properties}

Low-field susceptibility $4\pi \chi (T)$ data for good samples are shown in
Fig. \ref{lowfieldMT}. In this figure, the curves labeled $\chi _{{\rm zfc}}$
and $\chi _{{\rm fc}}$ were measured in the zero-field-cooled (zfc) and the
field-cooled (fc) states, respectively. The nominal superconducting volume
fraction was calculated from $f_{{\rm nom}}=-4\pi \chi _{{\rm zfc}}(T\ll
T_{c})$ and was not corrected for the demagnetization factors\cite
{Demagnetization}. The superconducting volume fractions were higher,
especially in the high magnetic field region than previous results\cite
{Jorgensen93,Er94,Er92,Er97,Tao92,Wiedenhorst98}.

The superconducting transition onset in Fig. \ref{lowfieldMT} appears at 43
K, which is the value typically reported for the La-112 compound\cite
{Ikeda93,Jorgensen93,Er94,Er92,Er91,Review,Yao94,Er97}. However, we can see
some notable differences from previous reports. One is a very sharp
transition near $T=43$ K, and another is a well-developed saturation of the
susceptibility at low temperatures, which reflects the formation of a
high-quality superconducting La-112 phase. The saturated values of $4\pi
\chi _{{\rm zfc}}$ at low temperatures are about $-1.0$, $-1.17$, and $-1.22$
for samples A, B, and C, respectively.

For a superconducting sphere with a radius $R$, $4\pi \chi _{{\rm zfc}}(T)$
is given by the Shoenberg formula\cite{Shoenberg} $-3/2(1-(3/x)\coth
x+3/x^{2})$, where $x=R/\lambda _{{\rm avg}}(T)$ and $\lambda _{{\rm avg}%
}(T) $ is the average magnetic penetration depth, {\em i.e.}, $\lambda _{%
{\rm avg}}=(\lambda _{ab}^{2}\lambda _{c})^{1/3}$. In the limit of $x\gg 1$,
the absolute value of $-4\pi \chi $ is not 1, but 1.5, due to the
demagnetization effect\cite{Demagnetization}. If we take the typical value
of $\lambda \sim 2000$ \AA ~for high-$T_{c}$ cuprates and the grain size $%
R\simeq 5~\mu $m obtained from the SEM image, the value of $4\pi \chi $ is
estimated to be about $-1.3$, which is close to the above measured value.
Thus, the real superconducting volume fraction of our sample should be close
to 100$\%$, especially for sample C. Our values were also confirmed using
fine-powdered samples, thus avoiding the possibility of weak links. Also the
zfc signal of the low-field magnetization $\chi _{{\rm zfc}}(T=5$ K$\ll
T_{c})$ was basically the same for 100 Oe as it was for 10 Oe, as can be
seen in Fig. \ref{lowfieldMT}, which was quite typical for all of our
samples, irrespective of the sample quality.

The irreversible field line, $H_{{\rm irr}}(T)$, from the high-field
magnetization up to 5 T showed that pinning was very strong in the
infinite-layer La-112 compound. In Fig. \ref{Hirr}, the magnetization curves
for fields higher than 1 T and the resulting $H_{{\rm irr}}(T)$ are
presented. The criterion for the reversible point was set as $|M_{{\rm zfc}%
}-M_{{\rm fc}}|=0.1$ emu/cm$^{3}$. The irreversible field was fitted with 
{\it H}$_{{\rm irr}}$({\it T})= {\it H}$_{0}$(1-{\it T}/{\it T}$_{c}$)$^{n}$%
. The best parameters were $H_{0}=55.7$ T, $T_{c}=42.6$ K, and $n=1.99$. Our
measured value of the irreversible field is about same order of magnitude as
previous results\cite{Er94}.

Magnetic hysteresis curves $M(H)$ were measured at temperatures between 5
and 30 K, as shown in Fig. \ref{MH} (a). The intragranular critical current
density was obtained from the relation $J_{c}\simeq 17(M_{\downarrow
}-M_{\uparrow })/R$,\cite{Clem87} where $M_{\uparrow }(M_{\downarrow })$ is
the magnetization in the increasing (decreasing) field branch in Gauss ($=$%
emu/cm$^{3}$) and $R$ $(\sim 5\mu m)$ is the average radius of the grains,
and is plotted in fig. \ref{MH} (b). Our value of $J_{c}$ is nearly one
order of magnitude larger than previous values\cite{Er94}.\ \ As an example, 
$J_{c}(5$ K$,$ $4$ T) $\approx $ $1.2\times 10^{6}$ A/cm$^{2}$ compared to
the previous value of $2\times 10^{5}$ A/cm$^{2}$.

The choice of $R$ was rather reasonable because the largest grains found in
the SEM images on many cleaved surfaces have $R_{{\rm max}}\sim 7.5$ $\mu $%
m, which guarantees the correct order of magnitude of our $J_{c}$\ value.
Also this $J_{c}$\ value was nearly same order of magnitude as that obtained
for powders using a sieve with the average size $R\simeq 3$ $\mu $m\cite
{MSKimunpublished}. The high-$T_{c}$ cuprate superconductors have strongly
2-dimensional characters, short coherence length, and `high' $T_{c}$. Due to
these, vortex lines become ill-defined and transform into pancake vortices
confined within the CuO$_{2}$ planes, which couple only weakly between the
layers. Thus the critical current density suffers great decrease at higher
temperatures due to the flux flow driven by a strong thermal fluctuation
effect. There are several extrinsic methods to enhance $J_{c}$. Correlated
defect was known to increase the pinning at higher temperatures and high
fields while point defects have been known to be efficient only at low
temperature and low fields\cite{Science}. The former such as columnar
defects generated by heavy-ion irradiation not only increases just pinning
centers but also could be thought to increase the coupling between vortices
along the irradiated trajectory. This is because the relatively strong
pinning centers are produced along straight line across CuO$_{2}$ plane. The
behavior of $J_{c}$ of La-112 here resembles the former case, namely the
critical current density does not decays fast as temperature and field
increases. For example, at $T$ $\sim T_{c}/2$, $J_{c}$ decreases by much
less than factor of 2 when the field is increased from 1 and 4.5 Tesla,
which is just typical behavior expected for Bean's critical state model\cite
{Bean}. Correlated defects are generally inserted into the sample on
purpose, surely absent in our samples. All these arguments suggest that the
behavior of $J_{c}$ of our sample is highly intrinsic because Jorgensen {\it %
et al}. showed that defects, most probably the oxygen defects and vacancies,
do not exist for this compound\cite{Jorgensen93,Er94,Er91}.

The samples studied here were made with in-situ annealing and showed that
the nearly uniform-sized grains were separated well from each other by wide
cracks. The uniform size made the superconducting transition sharp in the
low-field magnetization, and the cracks made the resistivity drop in the
transport measurement nearly invisible. Actually we tried to make samples
without in-situ annealing after sintering. This sample showed many smaller
grains between larger grains, which resulted in broad superconducting
transition in the low-field magnetization\ but with a clearer resistivity
drop due to a better connectivity between the grains. These suggest that the
use of grain radius not the sample radius is reasonable for the calculation
of $J_{c}$ like the previous report\cite{Er94}. The different $J_{c}$\
values between us and previous result\cite{Er94} seems to be partly due to
the uniformness of the grain size and/or the existence of many smaller
grains.

Now let's compare the above values with those of compounds having a charge
reservoir block, whose average distance between CuO$_{2}$ planes is larger.
Optimally doped (La,Sr)$_{2}$CuO$_{4}$ and (Nd,Ce)$_{2}$CuO$_{4}$ are the
most suitable for comparison with our electron-doped infinite-layer
superconductors because the former has nearly the same $T_{c}$\ as our
sample while the latter is an electron-doped cuprate superconductor similar
to ours.

First, $H_{{\rm irr}}(T/T_{c})$\ of La-112 is more than 2 times higher than
that of (La,Sr)$_{2}$CuO$_{4}$ and one order of magnitude larger than that
of Nd-214\cite{Er94,Fabrega92}. Similarly, our measured value of $J_{c}$ for
the La-112 compound is much higher than the value reported for
polycrystalline La-214, $J_{c}\simeq 1.7\times 10^{5}$ A/cm$^{2}$ at $4.2$\
K and $4$ T\cite{Senoussi87}. \ As for the Nd-214 compound, the reported $%
J_{c}\simeq 8\times 10^{5}$ A/cm$^{2}$ at $4.2$\ K and $0$ T was obtained
using a form of only a thin film not a bulk,\cite{Poole2000} so a direct
comparison is impossible. However, if the fact that the $J_{c}$\ of high-$%
T_{c}$ cuprates decreases by nearly one order of magnitude when the magnetic
field is increased from 0 to $\sim 5$ T is considered, the intragranular
critical current density of La-112 should be much larger than that of Nd-214 
\cite{Ginsberg}.

The above comparisons of $H_{{\rm irr}}(T)$\ and $J_{c}$ support the
conclusion that the interlayer coupling of an infinite-layer superconductor
is stronger due to the absence of a charge reservoir block. Such a stronger
interlayer coupling was also found with previous observation of the 3D
antiferromagnetic structure for an undoped infinite-layer compound, i.e., Ca$%
_{0.85}$Sr$_{0.15}$CuO$_{2}$. This material has been reported to have a
stronger 3-dimensional character than other parent insulators of cuprate
superconductors, such as YBa$_{2}$Cu$_{3}$O$_{6}$, La$_{2}$CuO$_{4}$, and Sr$%
_{2}$CuO$_{2}$Cl$_{2}$\cite{Lombardi96,Vaknin89,Keren93,Pozzi97}. For
example, an estimate of the ratio of the out-of-plane and the in-plane
coupling constants for Ca$_{0.85}$Sr$_{0.15}$CuO$_{2}$\ was two to three
orders of magnitude larger than corresponding values for YBa$_{2}$Cu$_{3}$O$%
_{6}$ and La$_{2}$CuO$_{4}$\cite{Lombardi96}. From our study, we claim that
the pinning properties of high-$T_{c}$ cuprates is improved at the extreme
limit of reducing the thickness of the charger reservoir block,\cite{Kim91}
i.e, at a cuprate superconductor without a charger reservoir block.

\section{summary}

We synthesized the infinite-layer compound Sr$_{0.9}$La$_{0.1}$CuO$_{2}$.
The quality of the samples was confirmed by using a structural analysis and
low-field magnetization measurements. Both the irreversibility field, $H_{%
{\rm irr}}(T)$,\ and the intragranular critical current density, $J_{c}$,
were found to be much higher than the values for (La,Sr)$_{2}$CuO$_{4}$ and
(Nd,Ce)$_{2}$CuO$_{4}$. And $J_{c}$\ does not decay fast as temperatrue and
magnetic field increases, unlike other cuprate superconductors. These
indicats an enhanced interlayer coupling between the CuO$_{2}$ planes\ due
to a shortening of the $c$-axis lattice constant.

\acknowledgments
We are thankful to K. Kadowaki, R. S. Liu, and D. Pavuna for useful
discussions on infinite-layer superconductors. We also greatly appreciate
our valuable discussions with P. D. Han, D. A. Payne, C. E. Lesher, M.
Takano, and A. Iyo on the general aspects of high-pressure synthesis. This
work is supported by the Ministry of Science and Technology of Korea through
the Creative Research Initiative Program.

\end{multicols}
\newpage

\begin{figure}[tbp]
\caption{Rietveld refinement of the X-ray powder diffraction pattern of
sample A. The dots are the raw data including background, and the solid line
is the calculated profile. The vertical tick marks below the profile
represent the positions of allowed diffractions in the tetragonal {\it P4/}%
mmm space group. A difference curve (observed pattern minus calculated
pattern) is also plotted at the bottom.}
\label{XRD}
\end{figure}

\begin{figure}[tbp]
\caption{Magnetic susceptibility, $4\protect\pi \protect\chi (T)$, of Sr$%
_{0.9}$La$_{0.1}$CuO$_{2}$\ for zero-field-cooling and field-cooling from
the low-field magnetization $M(T)$ at 10 and 100 Oe. For calculating the
nominal superconducting volume fraction, $f_{{\rm nom}}$, we used a
low-temperature density of $5.94$ g/cm$^{3}$ from Ref. 12. (a) Sample A, $f_{%
{\rm nom}}=100$\%, (b) Sample B, $f_{{\rm nom}}=117$\%, and (c) Sample C, $%
f_{{\rm nom}}=122$\%. }
\label{lowfieldMT}
\end{figure}

\begin{figure}[tbp]
\caption{$4\protect\pi M(T)$ curves of sample A at fields higher than 1
Tesla and irreversibility field {\it H}$_{irr}$($T$): (a) $4\protect\pi M(T)$
curves at 1, 2, 3, 4, and 5 Tesla and (b) irreversible field fitted with $%
H_{irr}(T)=H_{o}(1-T/T_{c})^{n}$. The criterion was chosen as $|M_{{\rm zfc}%
}-M_{{\rm fc}}|=0.1$ emu/cm$^{3}$. The uncertainty in terms of temperature
is less than 0.2 K.\ The fit was excellent with the parameters $H_{o}=55.7$
Tesla, $T_{c}=42.6$ K, and $n=1.99$. The top axis denotes the normalized
temperature $T$/$T_{c}$. The filled triangles were obtained with the
criterion $|M_{{\rm zfc}}-M_{{\rm fc}}|=0.01$ emu/cm$^{3}$.}
\label{Hirr}
\end{figure}

\begin{figure}[tbp]
\caption{(a) Magnetic hysteresis curves of sample A at 5, 10, 20, and 30 K.
(b) The field and the temperature dependences of the intragranular critical
current density $J_{c}$ were calculated by using $J_{c}\simeq
17(M_{\downarrow }-M_{\uparrow })/R$, where $M_{\uparrow }(M_{\downarrow })$
is the magnetization in the increasing (decreasing) field branch in Gauss ($%
= $emu/cm$^{3}$) and $R$ $(\sim 5\times 10^{-4}$ cm$)$ is the average radius
of the grains.}
\label{MH}
\end{figure}

\begin{table}[tbp]
\caption{Structural parameters for Sr$_{0.9}$La$_{0.1}$CuO$_{2}$ from
Rietveld refinement using the X-ray powder diffraction pattern for sample A.
The values in parentheses are reported ones based on the neutron powder
diffraction data in Ref 6.}
\label{refinement}
\begin{tabular}{ccc}
{Parameter} &  & Value \\ 
\tableline$a=b$ (\AA ) &  & 3.950 42 (3.950 68) \\ 
$c$ (\AA ) &  & 3.410 20 (3.409 02) \\ 
$V$ (\AA $^{3}$) &  & 53.219 (53.212) \\ 
$\alpha =\beta =\gamma $ &  & 90.000 0 \\ 
Sr, La\tablenotemark[1] & $x=y=z$ & 0.5 \\ 
& $n$ & 1 \\ 
Cu & $x=y=z$ & 0 \\ 
& $n$ & 1 \\ 
O & $x$ & 0.5 \\ 
& $y=z$ & 0 \\ 
& $n$ & 2 \\ 
\tableline Agreement factor &  & Value(\%) \\ 
\tableline R$_{wp}$ (\%) &  & 7.48(16.0) \\ 
R$_{p}$ (\%) &  & 4.68 \\ 
R$_{e}$ (\%) &  & 2.21 \\ 
\tableline Goodness of fit, $S$ &  & 3.3853
\end{tabular}
\tablenotetext[1]{Constraint: n(Sr):n(La)=0.9:0.1.}
\end{table}


\begin{thebibliography}{10}
\bibitem[*]{email}  Electronic address: jungking@postech.ac.kr

\bibitem{Siegrist88}  T. Siegrist, S. M. Zahurak, D. W. Murphy, and R. S.
Roth, Nature {\bf 334} (1988) 231.

\bibitem{Smith91}  M. G. Smith, A. Manthiran, J. Zhou, J. B. Goodenough, and
J. T. Markert, Nature {\bf 351} (1991) 549.

\bibitem{Sugii94}  Bobuyuki Sugii, H. Yamauchi, and Mitsuru Izumi, Phys.
Rev. B {\bf 50} (1994) 9503.

\bibitem{Jones94}  Edwin C. Jones, David P. Norton, David K. Christen, and
Douglas H. Lowndes, Phys. Rev. Lett. {\bf 73} (1994) 166.

\bibitem{Ikeda93}  N. Ikeda, Z. Hiroi, M. Azuma, M. Takano, and Y. Bando,
Physica C {\bf 210} (1993) 367.

\bibitem{Jorgensen93}  J. D. Jorgensen, P. G. Radaelli, D. G. Hinks, J. L.
Wagner, S. Kikkawa, G. Er, and F. Kanamaru, Phys. Rev. B {\bf 47} (1993) 14
654.

\bibitem{Er94}  P. Kobayashi, K. Kishio, B. Ni, K. Yamafuji, G. Er, F.
Kanamaru, S. Kikkawa, and M. Takano, Physica C {\bf 235} (1994) 2863.

\bibitem{Er92}  G. Er, S. Kikkawa, F. Kanamaru, Y. Miyamoto, S. Tanaka, M.
Sera, M. Sato, Z. Hiroi, M. Takano, and Y. Bando, Physica C {\bf 196} (1992)
271.

\bibitem{Er91}  G. Er, Y. Miyamoto, F. Kanamaru, and S. Kikkawa, Physica C 
{\bf 181} (1991) 206.

\bibitem{Review}  For a brief review of infinite-layer superconductors, see,
for example, J. T. Markert, K. Mochizuki, and A. V. Eliott, J. Low. Temp.
Phys. {\bf 105} (1996) 1367.

\bibitem{Yao94}  Xingjiang Zhou, Yushu Yao, Cheng Dong, Jingwei Li, Shunlian
Jia, and Zhongxian Zhao, Physica C {\bf 219} (1994) 123.

\bibitem{Er97}  G. Er, S. Kikkawa, M. Takahashi, F. Kanamaru, M. Hangyo, K.
Kisoda, and S. Nakashima, Physica C {\bf 290} (1997) 1.

\bibitem{Kim91}  D. H. Kim, K. E. Gray, R. T. Kampwirth, J. C. Smith, D. S.
Richeson, T. J. Marks, J. H. Wang, J.\ Tallvacchio, and M. Eddy, Physica C 
{\bf 177} (1991) 431.

\bibitem{Senoussi87}  S. Senoussi, M. Oussena, M. Pribault, and G. Collin,
Phys. Rev. B {\bf 36} (1987) 4003.

\bibitem{Izumi}  K. Kinoshita, F. Izumi, T. Yamada, and H. Asano, Phys. Rev.
B {\bf 45} (1992) 5558.

\bibitem{Demagnetization}  We should be cautious with polycrystalline
samples because the microscopic ($\mu m$) shape of each grain and its angle
with the external magnetic field, not the macroscopic({\it mm} scale) shape
of the sample, is important in selecting the actual demagnetization factor.

\bibitem{Tao92}  S. Tao, H.-U. Nissen, C. Beeli, M. Cantoni, M. G. Smith, J.
Zhou, and J. B. Goodenough, Physica C {\bf 204} (1992) 117.

\bibitem{Wiedenhorst98}  B. Wiedenhorst, H. Berg, R. Gross, B. H. Freitag,
and W. Mader, Physica C {\bf 304} (1998) 147.

\bibitem{Shoenberg}  D. Shoenberg, Superconductivity (Cambridge University,
Cambridge, 1954), p. 164.

\bibitem{Clem87}  John R. Clem and V. G. Kogan, Jpn. J. Appl. Phys. {\bf 26}
(1987) 1162.

\bibitem{MSKimunpublished}  Mun-Seog Kim, C. U. Jung, {\it et al}.,
unpublished.

\bibitem{Science}  I. Chong, Z. Hiroi, M. Izumi, J. Shimoyama, Y. Nakayama,
K. Kishio, T. Terashima, Y. Bando, and M. Takano, Science, {\bf 276}, (1997)
770 and references therein.

\bibitem{Bean}  C, .~P. Bean, Rev. Mod. Phys. {\bf 36}, (1964) 31.

\bibitem{Fabrega92}  L. F\'{a}brega, J. Fontcuberta, S. Pi\~{n}ol, C. J. van
der Beek, and P. H. Kes, Phys. Rev. B {\bf 46} (1992) 11 952.

\bibitem{Poole2000}  Charles P. Poole, Handbook of Superconductivity
(Academic Press, San Diego, 2000), p. 466.

\bibitem{Ginsberg}  D. M. Ginsberg, Physical Properties of High Temperature
Superconductors I (World Scientific, Singapore, 1989), p. 278.

\bibitem{Lombardi96}  A. Lombardi, M. Mali, J. Roos, and D. Brinkmann, Phys.
Rev. B {\bf 54} (1996) 93.

\bibitem{Vaknin89}  D. Vaknin, E. Caignon, P. K. Davies, and J. E. Fischer,
D. C. Johnston, and D. P. Goshorn, Phys. Rev. B {\bf 39} (1989) 9122.

\bibitem{Keren93}  A. Keren, L. P. Le, G. M. Luke, B. J. Sternlieb, W. D.
Wu, Y. J. Uemura, S. Tajima, and S. Uchida, Phys. Rev. B {\bf 48} (1993) 12
926.

\bibitem{Pozzi97}  R. Pizzi, M. Mali, M. Matsumura, F. Raffa, J. Roos, and
D. Brinkmann, Phys. Rev. B {\bf 56} (1997) 759.

\end{thebibliography}
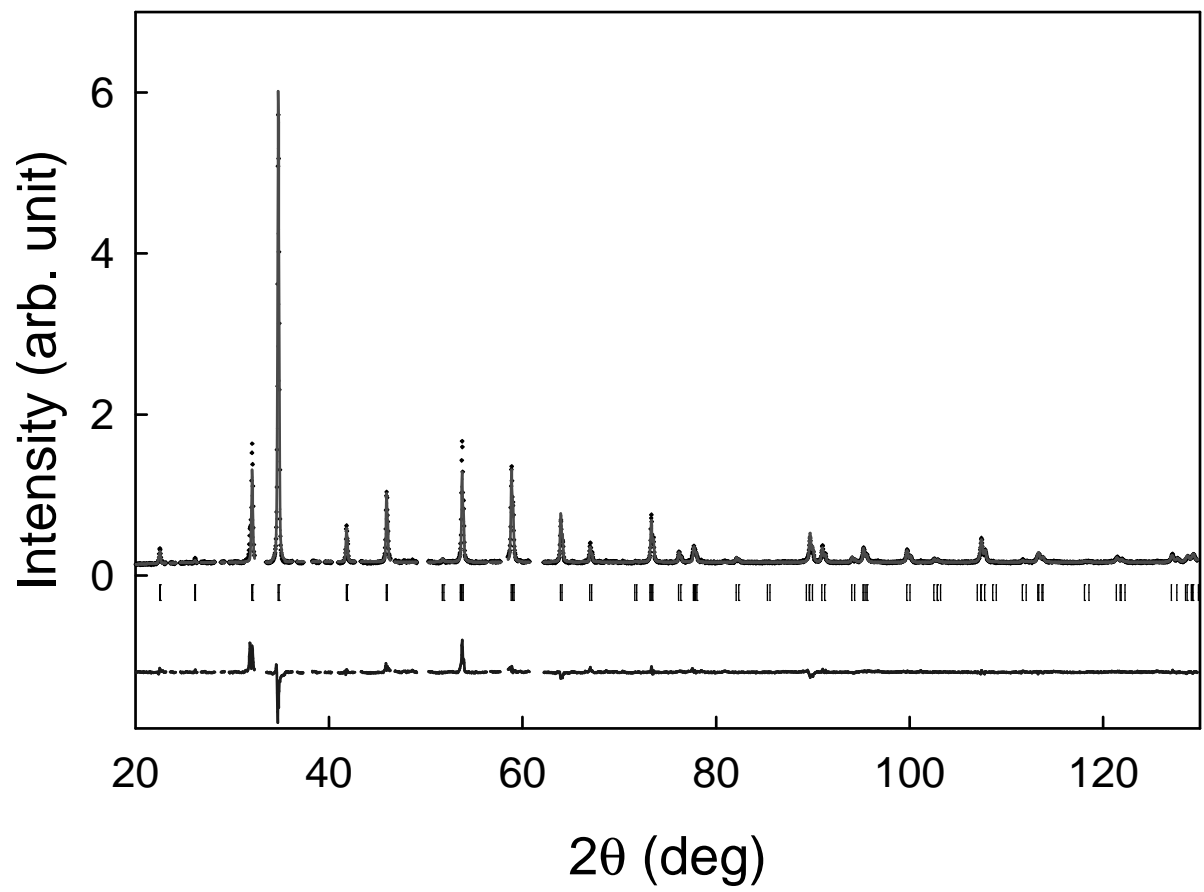
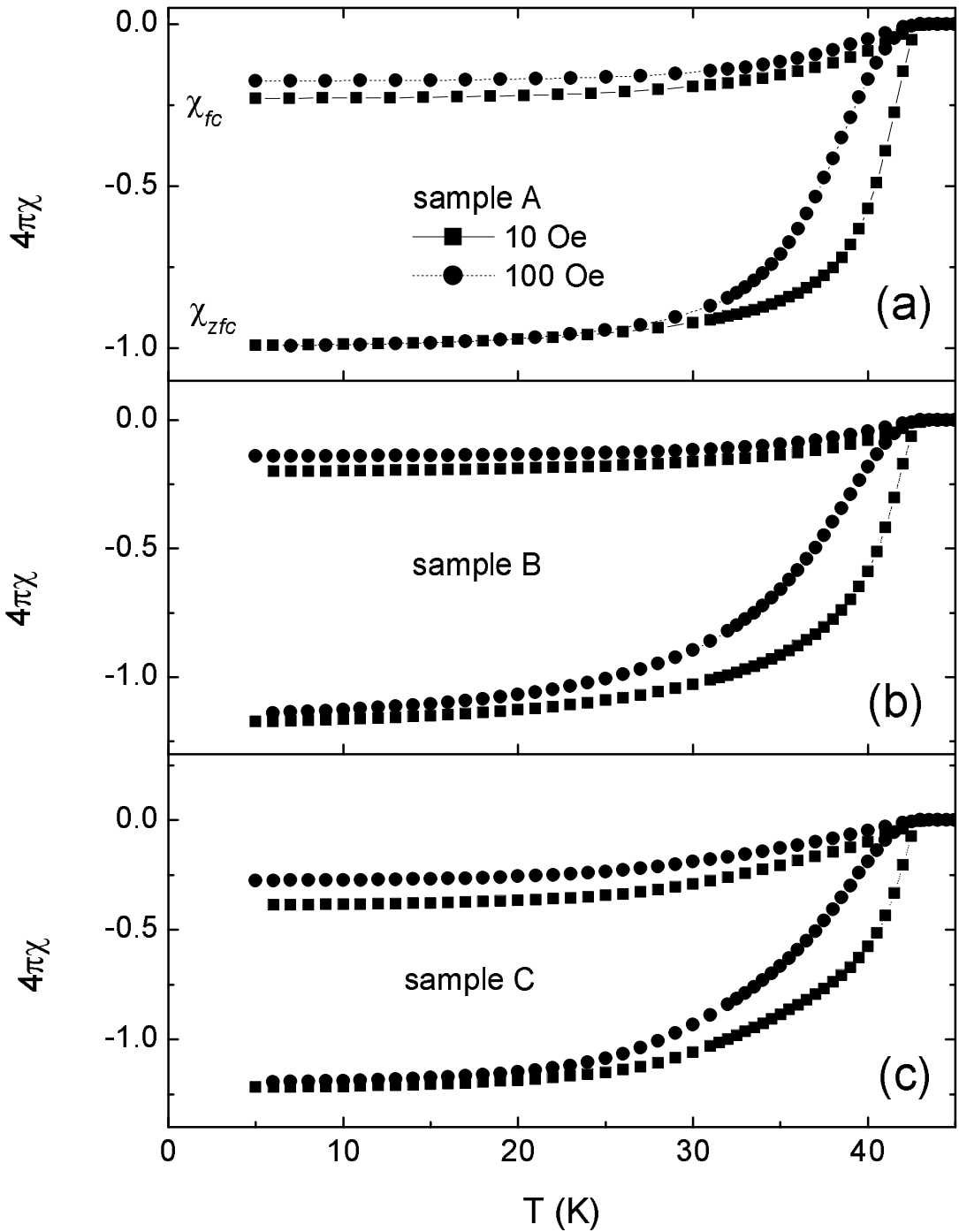
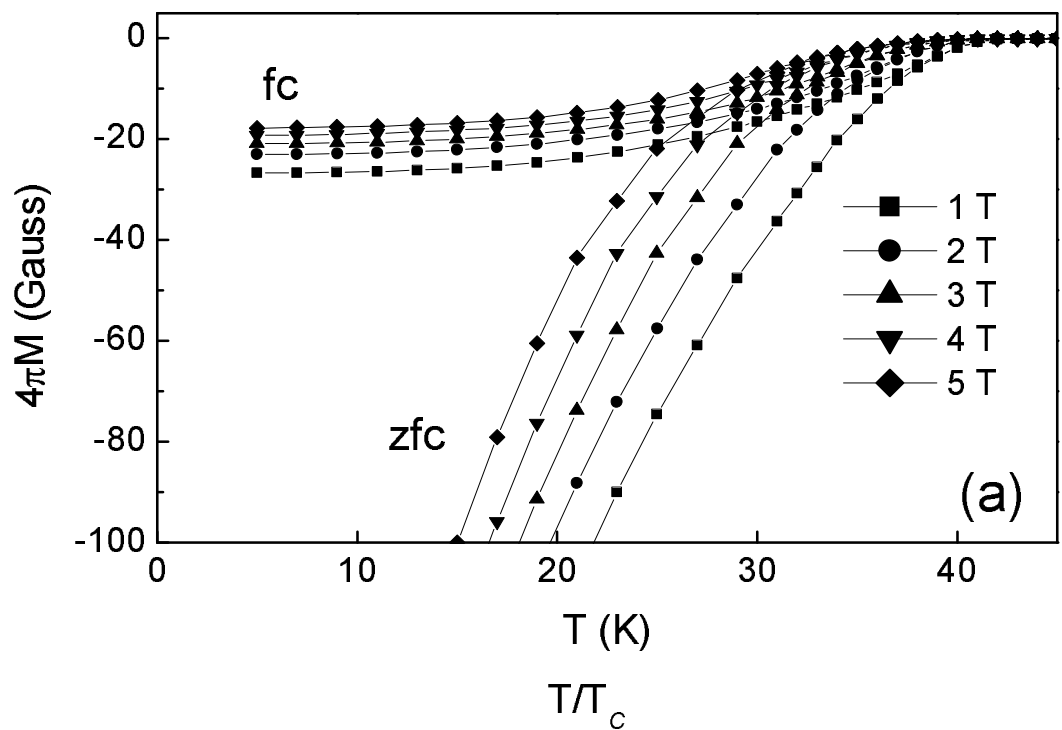
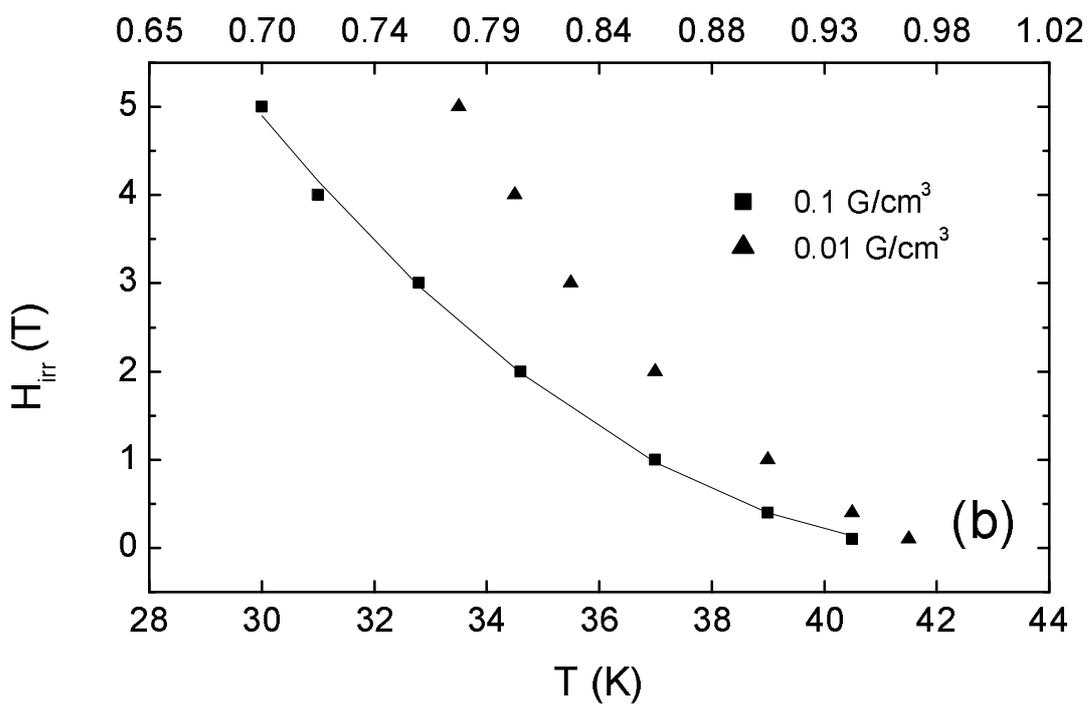
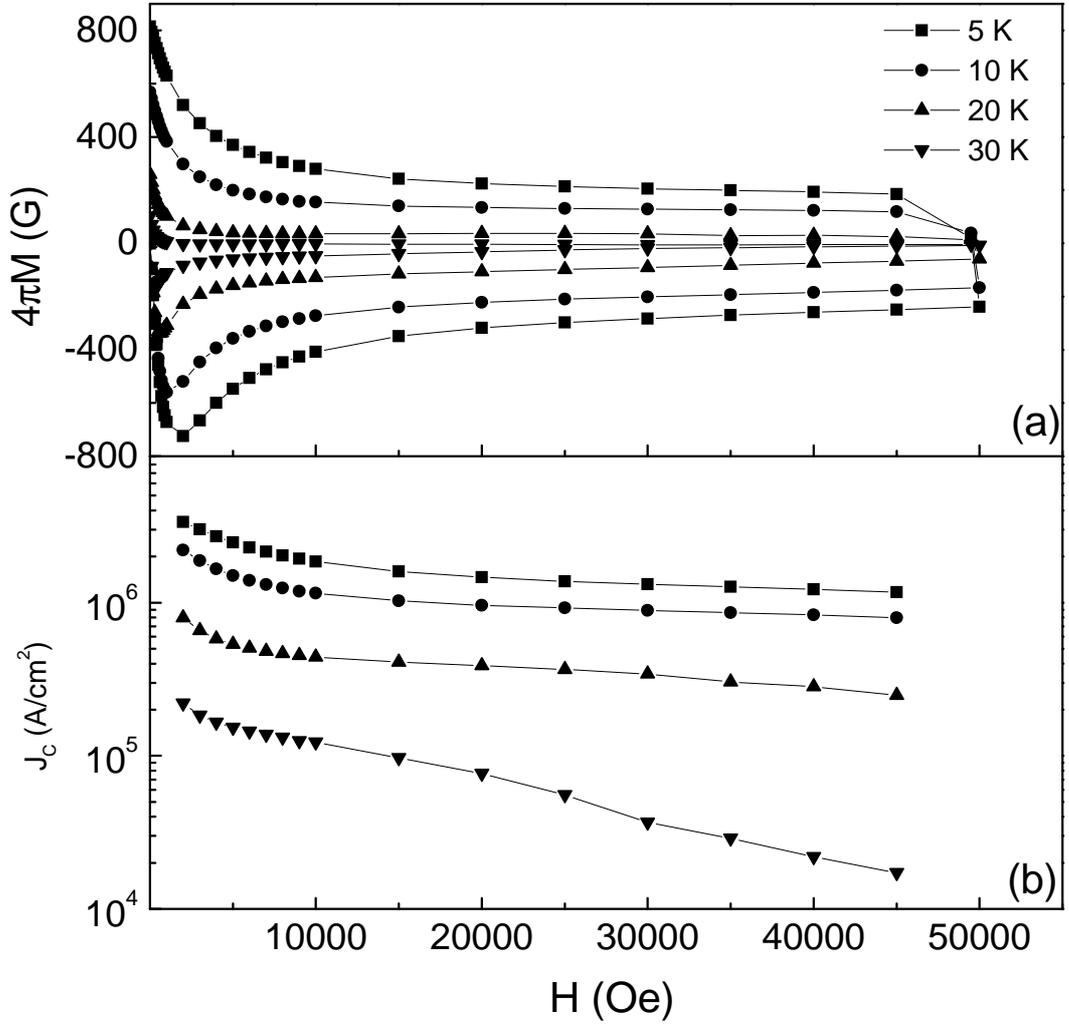
\end{document}